\documentclass{article}
\usepackage{heron2e}
\title{Representations of Two-Colour BWM Algebras and 
       Solvable Lattice Models}
\author{Uwe Grimm}
\address{Institut f\"{u}r Physik,\\
         Technische Universit\"{a}t Chemnitz,\\
         D--09107 Chemnitz, Germany}

\usepackage{amssymb}

\newcommand{\unitmat}{\mathbb{I}}
\newcommand{\complexnum}{\mathbb{C}}

\newcommand{\ppp}[2]{\; p^{(#1,#2)}}
\newcommand{\bbb}[2]{\; b^{(#1,#2)}}
\newcommand{\bbbp}[2]{
\; b\mbox{\raisebox{.68em}{$\!\scriptscriptstyle +$}}{}^{(#1,#2)}}
\newcommand{\bbbm}[2]{
\; b\mbox{\raisebox{.68em}{$\!\scriptscriptstyle -$}}{}^{(#1,#2)}}
\newcommand{\bbbpm}[2]{
\; b\mbox{\raisebox{.68em}{$\!\scriptscriptstyle\pm$}}{}^{(#1,#2)}}
\newcommand{\eee}[2]{\; e^{(#1,#2)}}

\newcommand{\pppj}[2]{\ppp{#1}{#2}_{j}}
\newcommand{\bbbj}[2]{\bbb{#1}{#2}_{j}}
\newcommand{\bbbpj}[2]{\bbbp{#1}{#2}_{\!\! j}}
\newcommand{\bbbmj}[2]{\bbbm{#1}{#2}_{\!\! j}}
\newcommand{\bbbpmj}[2]{\bbbpm{#1}{#2}_{\!\! j}}
\newcommand{\eeej}[2]{\eee{#1}{#2}_{j}}

\begin{document}
\maketitle
\begin{abstract}
Many of the known solutions of the Yang-Baxter equation, 
which are related to solvable lattice models of vertex- and IRF-type, 
yield representations of the Birman-Wenzl-Murakami algebra. 
{}From these, representations of a two-colour generalization of the 
Birman-Wenzl-Murakami algebra can be constructed, which in turn are used 
to derive trigonometric solutions to the Yang-Baxter equation. In spirit, this 
construction resembles the fusion procedure, in the sense that starting 
from known solutions of the Yang-Baxter equation new solutions can be 
obtained.
\end{abstract}

\section{Introduction}

Over the the past fifteen years, since Baxter's famous book first appeared 
in print \cite{Baxter}, the full relevance of the Yang-Baxter equation (YBE) 
in the theory of solvable two-dimensional lattice models has been 
realized. Moreover, many fruitful and partly unexpected connections
to other branches of mathematics and physics, including for example 
quantum groups, knot and link invariants, and integrable quantum 
field theories with purely elastic (factorized) scattering, have 
been discovered.

Among the several algebraic techniques used to construct solutions to
the YBE, a particularly interesting approach is based on braid-monoid
algebras (BMA) \cite{WadDegAku}. By a procedure called 
{\em Baxterization} \cite{Jones}, one reduces the problem of
finding representations of the  Yang-Baxter algebra (YBA) to 
the simpler task of constructing representations of a certain
BMA. Examples of BMA for which a Baxterization is known 
are the Temperley-Lieb algebra \cite{TempLieb},
the Birman-Wenzl-Murakami (BWM) algebra \cite{BirWen,Mura} and their 
recently introduced \cite{GriPea93}
`dilute' \cite{Roche,WarNieSea,Gri94a,Gri94b,War95,GriWar95a} 
and `two-colour' generalizations \cite{WarNie93,GriWar95b}.

In this short note, we present explicit expressions for 
`vertex-type' representations of the two-colour BWM (TCBWM) algebra,
by which we understand representations acting on a tensor product space.
This is complementary to the content of Ref.~\cite{GriWar95b}
where exclusively RSOS (restricted solid-on-solid) type representations
were considered, for which the representation space is spanned
by the set of all paths on certain graphs.
Here, the representations are constructed by suitably combining 
representations of the ordinary BWM algebra underlying the solvable 
vertex models of Bazhanov \cite{Bazh} and Jimbo \cite{Jimbo}. By
the known Baxterization of the TCBWM algebra \cite{GriWar95b},
this yields solutions of the YBE and hence solvable lattice models
of vertex type.

In contrast to the case of RSOS type representations discussed 
in Ref.~\cite{GriWar95b}, we do not expect that
the corresponding solvable vertex models are new. Instead,
we expect a similar scenario as in the dilute case, where the $R$-matrices
constructed from representations of the dilute BWM algebra are related
by a suitable gauge transformation to other members of the lists of
well-known $R$-matrices \cite{Bazh,Jimbo}. For one series of 
$R$-matrices, this has been explicitly demonstrated in Ref.~\cite{Gri94a}.

\section{The TCBWM algebra}

For lack of space, we cannot repeat the complete definition of the
TCBWM algebra here, thus we only give a rough sketch of the idea,
see Refs.~\cite{GriPea93,GriWar95b} for details.

An ordinary BMA \cite{WadDegAku} is generated by two types of 
generators, the braids (with their inverses) and 
the monoids or Temperley-Lieb (TL) operators. Besides the usual braid
relations and the defining relations of the TL algebra, a number of
relations involving both types of generators are imposed. 
All relations have a natural interpretation in terms of continuous moves
in a diagrammatic presentation of the generators acting on arrays
of strings \cite{WadDegAku}, hinting at the connection to the theory of
knot and link invariants. 
The defining relations also involve two central elements 
customarily denoted by $\sqrt{Q}$ and $\omega$, which in the graphical
interpretation are associated to closed loops and so-called `twists',
respectively. The BWM algebra is a certain quotient of a BMA, in which 
the braid generators satisfy a cubic reduction relation \cite{BirWen,Mura}.

The TCBWM algebra is a two-colour generalization of the BWM algebra, with
two species (colours) of strings. The generators now carry colour indices
\begin{eqnarray}
\pppj{c}{c'} & & \mbox{(projectors)}\nonumber \\*
\bbbpmj{c}{c}\; ,\;\bbbj{c}{\bar{c}} & & \mbox{(braids)}\nonumber \\* 
\eeej{c}{c'} & & \mbox{(Temperley-Lieb operators)} 
\label{gen}
\end{eqnarray}
and one has additional generators $\ppp{c}{c'}_j$
which act as projection operators
on certain colours. Here, $c,c'\in\{1,2\}$ label the two colours,
and $\bar{c}=3-c$ denotes the complementary (or opposite) colour. 
Furthermore, the generators carry a label $j\in\{1,2,\ldots,N-1\}$. In the
graphical interpretation, this means that the corresponding generator
only acts on the two of the $N$ strings, namely on those labeled 
by $j$ and $j+1$. The central elements mentioned above 
now also carry colour indices, thus we have factors of $\sqrt{Q_c}$ 
and $\omega_c^{}$ associated to closed loops and to
`twists' in a string of colour $c$, respectively.

The defining relations of the TCBWM algebra can most easily be 
understood directly from the diagrammatic interpretation: they are 
given by considering all possible colourings of the diagrams corresponding
to the defining relations of the ordinary BWM algebra. In addition, 
we demand that any products leading to diagrams with colour mismatches 
vanish in the algebra. This statement basically reduces
to the requirement that the generators $\ppp{c}{c'}_j$ are indeed
orthogonal projectors.

Before we move on to the representations, let us briefly present
the relations defining the BWM quotient. 
For the coloured braids, one has the cubic reduction relations
\begin{equation}
\left(\bbbpj{c}{c} - q_c^{-1} \pppj{c}{c}\right) 
\left(\bbbpj{c}{c} + q_c^{} \pppj{c}{c}\right) 
\left(\bbbpj{c}{c} - \omega_c^{} \pppj{c}{c}\right) 
\;\; =\;\; 0 \; ;
\label{cubic}
\end{equation}
and the Temperley-Lieb generators $\eeej{c}{c}$ are given by 
quadratic expressions in the braids as follows
\begin{equation}
\eeej{c}{c}\;\; = \;\; \pppj{c}{c}\; +\; 
\frac{\bbbpj{c}{c}-\bbbmj{c}{c}}{q_c^{}-q_c^{-1}}\; ,
\label{quad}
\end{equation}
where $q_c^{}$ is related to $\sqrt{Q_c}$ and $\omega_c^{}$ by
\begin{equation}
\sqrt{Q_c} \;\; = \;\; 1\; +
\;\frac{\omega_c^{}-\omega_{c}^{-1}}{q_c^{}-q_c^{-1}}
\label{Q}
\end{equation}
which is a consequence of Eqs.~(\ref{cubic}) and (\ref{quad}) and
the defining relations of the TL algebra.

\section{Representations of the TCBWM algebra}

We now present explicit expressions for representations
of the TCBWM algebra. We build these representations on
a pair of representations of the ordinary BWM algebra,
which are labeled by affine Lie algebras $\mbox{B}^{(1)}_n$,
$\mbox{C}^{(1)}_n$, and $\mbox{D}^{(1)}_n$ \cite{WadDegAku} and
underlie the corresponding series of $R$-matrices of 
Refs.~\cite{Bazh,Jimbo}. Thus the TCBWM representations are labeled by 
pairs $(\mbox{G}_{1,n_1}^{(1)},\mbox{G}_{2,n_2}^{(1)})$,
where $\mbox{G}_c\in\{\mbox{B},\mbox{C},\mbox{D}\}$ and the index 
$c$ refers to the colour. Given $\mbox{G}_c$ and $n_c$, the
representation still contains two parameters $q_c^{}$, which
determine $\omega_c^{}$ through
\begin{equation}
\omega_c^{}\;\; =\;\;\left\{\begin{array}{@{\,}l@{\qquad}l@{}}
\hphantom{-}\, q_c^{2n_c} & \mbox{if $\mbox{G}_c=\mbox{B}$}\\
-\, q_c^{2n_c+1} & \mbox{if $\mbox{G}_c=\mbox{C}$}\\
\hphantom{-}\, q_c^{2n_c-1} & \mbox{if $\mbox{G}_c=\mbox{D}$}
\end{array}\right.
\label{omega}
\end{equation}
and thereby $\sqrt{Q_c}$ by Eq.~(\ref{Q}).

The two-colour BWM representations we are interested in act 
on the tensor product space
\mbox{$V=\bigotimes_{j=1}^{N}\complexnum^{d}_{}$}, 
where $d=d_1+d_2$ is given by
\begin{equation}
d_c\;\; =\;\;\left\{\begin{array}{@{\,}l@{\qquad}l@{}}
2n_c+1 & \mbox{if $\mbox{G}_c=\mbox{B}$}\\
2n_c & \mbox{if $\mbox{G}_c\in\{\mbox{C},\mbox{D}\}$}
\end{array}\right.
\end{equation}
and \mbox{$V_c=\bigotimes_{j=1}^{N}\complexnum^{d_c}_{}$} is the
representation space for the single-colour representations.
In these representations, the generators (\ref{gen}) act 
non-trivially only in two factors (labeled by $j$ and $j+1$) 
of the $N$-fold tensor product, wherefore it suffices to give 
the expressions for two factors which thus are $d^2\times d^2$ matrices
which we denote by the same symbols (\ref{gen}) 
without the site index $j$.

We introduce two index sets
\begin{eqnarray}
I_1 & = & \{1,2,\ldots,d_1\} \\
I_2 & = & \{d_1+1,d_1+2,\ldots,d_1+d_2\}
\end{eqnarray}
to keep apart labels referring to different colours.
For $\alpha\in I_c$, we define
\begin{eqnarray}
\tilde{\alpha} & = & \alpha - \frac{d_c+1}{2} - d_1\, \delta_{c,2}
\;\;\in \;\; 
\left\{-\frac{d_c-1}{2},-\frac{d_c-3}{2},\ldots,
\frac{d_c-1}{2}\right\} \\
\alpha' & = & d_c + 1 - \alpha + 2 d_1 \delta_{c,2}\;\; \in\;\; I_c
\end{eqnarray}
such that the possible values of $\tilde{\alpha}$ for both colours 
lie symmetrically around zero. 
Furthermore, we introduce symbols $\varepsilon_{\alpha}^{}$ 
and $\bar{\alpha}$ which are defined as follows:
\begin{eqnarray}
\mbox{For $\alpha\in I_c$ and $\mbox{G}_c\in\{\mbox{B},\mbox{D}\}$:
\hphantom{$=\mbox{C}$}}\!\!\!\!\!\!\!\!\!\!
& & \varepsilon_{\alpha}^{}\hphantom{\bar{\alpha}}\! =\;\; 1 \\
& & \bar{\alpha}\hphantom{\varepsilon_{\alpha}^{}}\! =\;\; \left\{
\begin{array}{@{\,}l@{\qquad}l@{}}
\tilde{\alpha} + 1/2 & \alpha<\alpha' \\ 
\tilde{\alpha}       & \alpha=\alpha' \\ 
\tilde{\alpha} - 1/2 & \alpha>\alpha' 
\end{array}\right.\qquad \\
\mbox{For $\alpha\in I_c$ and $\mbox{G}_c=\mbox{C}$:
\hphantom{$\in\{\mbox{B},\mbox{D}\}$}}\!\!\!\!\!\!\!\!\!\!
& & \varepsilon_{\alpha}^{}\hphantom{\bar{\alpha}}\! =\;\; \left\{
\begin{array}{@{\,}r@{\qquad}l@{}}
 1 & \alpha<\alpha' \\ 
-1 & \alpha>\alpha' 
\end{array}\right.\qquad \\
& & \bar{\alpha}\hphantom{\varepsilon_{\alpha}^{}}\! =\;\; \left\{
\begin{array}{@{\,}l@{\qquad}l@{}}
\tilde{\alpha} - 1/2 & \alpha<\alpha' \\ 
\tilde{\alpha} + 1/2 & \alpha>\alpha' 
\end{array}\right.\qquad
\end{eqnarray}
Note that $\alpha=\alpha'\in I_c$ can only occur for representations
labeled by $B^{(1)}_{c,n_c}$ because these have an odd $d_c=2n_c+1$.

With these notations, we obtain a two-colour representation by setting
\begin{eqnarray}
\ppp{c}{c'} & = & 
\sum_{\alpha\in I_c}\; 
\sum_{\beta\in I_{c'}}\; 
E_{\alpha,\alpha} \otimes E_{\beta,\beta} \\
\bbbp{c}{c} & = & 
\sum_{\alpha\in I_c}\; q_c^{-1}\, 
[1+(q_c^{}-1)\,\delta_{\alpha,\alpha'}]\;
E_{\alpha,\alpha} \otimes E_{\alpha,\alpha} \nonumber \\*
& & \; +\;
\sum_{\alpha\ne\beta\in I_c}\;
[1+(q_c^{}-1)\,\delta_{\alpha,\beta'}]\;
E_{\alpha,\beta} \otimes E_{\beta,\alpha} \nonumber \\*
& & \; -\; (q_c^{}-q_c^{-1})\; 
\sum_{\alpha<\beta\in I_c}\; 
E_{\alpha,\alpha} \otimes E_{\beta,\beta} \nonumber \\*
& & \; +\; (q_c^{}-q_c^{-1})\; 
\sum_{\alpha>\beta\in I_c}\; 
\varepsilon_{\alpha}\,\varepsilon_{\beta}\, 
q_c^{\bar{\alpha}-\bar{\beta}}\; 
E_{\alpha',\beta} \otimes E_{\alpha,\beta'}\\
\bbbm{c}{c} & = & 
\sum_{\alpha\in I_c}\; q_c^{}\, 
[1+(q_c^{-1}-1)\,\delta_{\alpha,\alpha'}]\;
E_{\alpha,\alpha} \otimes E_{\alpha,\alpha} \nonumber \\*
& & \; +\; \sum_{\alpha\ne\beta\in I_c}\; 
[1+(q_c^{-1}-1)\,\delta_{\alpha,\beta'}]\;
E_{\alpha,\beta} \otimes E_{\beta,\alpha} \nonumber \\*
& & \; -\; (q_c^{}-q_c^{-1})\; 
\sum_{\alpha>\beta\in I_c}\; 
E_{\alpha,\alpha} \otimes E_{\beta,\beta} \nonumber \\*
& & \; -\; (q_c^{}-q_c^{-1})\; 
\sum_{\alpha<\beta\in I_c}\; 
\varepsilon_{\alpha}\,\varepsilon_{\beta}\, 
q_c^{\bar{\alpha}-\bar{\beta}}\; 
E_{\alpha',\beta} \otimes E_{\alpha,\beta'}\\
\bbb{c}{\bar{c}} & = & 
\sum_{\alpha\in I_c}\;
\sum_{\beta\in I_{\bar{c}}}\; 
E_{\beta,\alpha} \otimes E_{\alpha,\beta}\\
\eee{c}{\bar{c}} & = & 
\sum_{\alpha\in I_c}\; 
\sum_{\beta\in I_{\bar{c}}}\; 
\varepsilon_{\alpha}^{}\,\varepsilon_{\beta}^{}\, 
q_c^{\bar{\alpha}}\, q_{\bar c}^{-\bar{\beta}}\;
E_{\alpha',\beta} \otimes E_{\alpha,\beta'}
\end{eqnarray}
where, as above, $c,c'\in\{1,2\}$ and $\bar{c}=3-c$. The $d\times d$ 
matrices $E_{\alpha,\beta}$ have matrix elements
$(E_{\alpha,\beta})_{i,j}=\delta_{i,\alpha}\delta_{j,\beta}$.
The expressions for the generators $\eee{c}{c}$ follow from Eq.~(\ref{quad}).

\section{Solvable vertex models}

The Boltzmann weights of solvable vertex models derived from
representations of the TCBWM algebra are encoded in the matrix 
elements of the $R$-matrix.
The Baxterization of Ref.~\cite{GriWar95b} yields a general 
expression for a trigonometric $R$-matrix for {\em any\/} 
representation given in the previous section, 
with the restriction that the parameters $q_c$ have to be equal,
i.e., $q_1=q_2=q$. Let us introduce two parameters $\lambda$ and $\eta$ by
\begin{equation}
q\;\; = \;\;\exp(-i\lambda)\; , \qquad
q^{2\eta}_{} \;\; = \;\; \exp(-2i\eta\lambda)
\;\; = \;\; \omega_{1}^{}\,\omega_{2}^{}\; ,
\end{equation}
where of course the values of $\omega_c^{}$ are determined by 
Eq.~(\ref{omega}).

Denoting the spectral parameter by $u$, the $R$-matrix has the 
form \cite{GriWar95b}
\begin{eqnarray}
\check{R}(u)\;\; = & \displaystyle\left.\rule[0em]{0em}{2em}\!\!
\sum_{c=1}^{2}\;\;\right\{ \!\!\!\!\!\! &
\ppp{c}{c}
\;+\;\frac{\sin(\eta\lambda-u)}{\sin(\eta\lambda)}\; \ppp{c}{\bar{c}}
\nonumber\\*
& &\left.\rule[0em]{0em}{2em}
\; -\;\frac{\sin(u)}{2i\,\sin(\lambda)\,\sin(\eta\lambda)}\:\left(\,
       e^{i(\eta\lambda-u)}\, \bbbp{c}{c}\: -\:
       e^{-i(\eta\lambda-u)}\, \bbbm{c}{c}\, \right)\right. \nonumber\\*
& &\left.\rule[0em]{0em}{2em}\; +\;
    \frac{\sin(u)\,\sin(\eta\lambda-u)}{\sin(\lambda)\,\sin(\eta\lambda)}
    \;\bbb{c}{\bar{c}}
    \;+\;\frac{\sin(u)}{\sin(\eta\lambda)}\;\eee{c}{\bar{c}}\;\right\} 
\label{R}
\end{eqnarray}
which is completely symmetric in the two colours. It satisfies the 
quantum YBE
\begin{equation}
\left(\check{R}(u)\otimes\unitmat\right)
\left(\unitmat\otimes\check{R}(u+v)\right)
\left(\check{R}(v)\otimes\unitmat\right) \;\; = \;\;
\left(\unitmat\otimes\check{R}(v)\right)
\left(\check{R}(u+v)\otimes\unitmat\right)
\left(\unitmat\otimes\check{R}(u)\right)
\end{equation}
where $\unitmat$ denotes the $d\times d$ unit matrix.
These solutions of the YBE are crossing-symmetric \cite{WadDegAku}
with crossing parameter $\eta\lambda$, and satisfy the inversion relation
\begin{equation} 
\check{R}(u)\,\check{R}(-u) \;\; = \;\; \varrho(u)\, \varrho(-u)\;
\unitmat\otimes\unitmat
\end{equation}
where the function $\varrho(u)$ is given by
\begin{equation}
\varrho(u) \;\; = \;\; 
\frac{\sin(\lambda-u)\,\sin(\eta\lambda-u)}{\sin\lambda\,\sin\eta\lambda}\; .
\end{equation}

\section{Concluding remarks}

We presented explicit representations of the TCBWM algebra
acting on tensor product spaces. These representations are
built on a pair of two representations of the ordinary BWM algebra.
{}From any representation of this type, we construct an $R$-matrix
via the Baxterization derived in Ref.~\cite{GriWar95b}, and hence
a solvable lattice model of vertex type. Although the construction 
shows some similarity to the fusion procedure, these models 
are different from those constructed
by fusion, compare e.g.\ Ref.~\cite{ZhouPeaGri} for the 
fusion of the dilute $\mbox{A}_L$ models.
Note that the $R$-matrices related to the dilute BWM 
algebra \cite{Gri94a,Gri94b} and to the dilute and 
two-colour TL algebras \cite{WarNie93,GriPea93} can be recovered
as special cases of the TCBWM result, see Ref.~\cite{GriWar95b}.

As already mentioned in the introduction, we expect that the 
$R$-matrices obtained in this way are not new, but related to
other trigonometric $R$-matrices within the lists given in 
Refs.~\cite{Bazh,Jimbo} by gauge transformations. It would be 
interesting to know the precise relationship between these 
$R$-matrices and to understand their construction on the level 
of the underlying quantum groups.

\end{document}